%% file: maininit.tex
\documentclass[ALICE,manyauthors]{cernphprep}
\usepackage[comma,square,numbers,sort&compress]{natbib}

\usepackage{lineno}
\usepackage{xspace}
\usepackage{hyperref}
\usepackage{xcolor}
\usepackage[T1]{fontenc}
\usepackage{orcidlink}

\begin{document}
\input{commands.tex}
%\linenumbers
%%%%%%%%%%%%%%%  Title page %%%%%%%%%%%%%%%%%%%%%%%%
\begin{titlepage}
% the dates below correspond to CERN approval
% please don't touch: EB chairs will take care
\PHyear{2022}       % required, will be obtained from CERN
\PHnumber{104}      % required, will be obtained from CERN
\PHdate{19 May}  % required, will be obtained from CERN
%%%%%%%%%%%%%%%%%%%%%%%%%%%%%%%%%%%%%%%%%%%%%%%%%%%%

%%% Put your own title + short title here:
\title{$\Sigma(1385)^{\pm}$ resonance production in \PbPb collisions at \fivenn}
\ShortTitle{$\Sigma(1385)^{\pm}$  in \PbPb}   % appears on left page headers

%%% Do not change the next lines
\Collaboration{ALICE Collaboration\thanks{See Appendix~\ref{app:collab} for the list of collaboration members}}
\ShortAuthor{ALICE Collaboration} % appears on right page headers, do not change

\begin{abstract}
Hadronic resonances are used to probe the hadron gas produced in the late stage of heavy-ion collisions since they decay on the same timescale, of the order of 1 to 10 fm/$c$, as the decoupling time of the system.
In the hadron gas, (pseudo)elastic scatterings among the products of resonances that decayed before the kinetic freeze-out and regeneration processes counteract each other, the net effect depending on the resonance lifetime, the duration of the hadronic phase, and the hadronic cross sections at play.
In this context, the $\Sigma(1385)^{\pm}$ particle is of particular interest as models predict that regeneration dominates over rescattering despite its relatively short lifetime of about 5.5 fm/$c$. 
The first measurement of the $\Sigma(1385)^{\pm}$ resonance production at midrapidity in \PbPb  collisions at $\sqrt{s_{\mathrm{NN}}}= 5.02$ TeV with the ALICE detector is presented in this Letter. The resonances are reconstructed via their hadronic decay channel, $\Lambda\pi$, as a function of the transverse momentum ($p_{\rm T}$) and the collision centrality.
The results are discussed in comparison with the measured yield of pions and with expectations from the statistical hadronization model as well as commonly employed event generators, including \Angantyr and EPOS3 coupled to the UrQMD hadronic cascade afterburner. None of the models can describe the data. For $\Sigma(1385)^{\pm}$, a similar behaviour as {\ensuremath{{\rm K}^{*} (892)^{0}}\xspace} is observed in data unlike the predictions of  EPOS3 with afterburner.

\end{abstract}
\end{titlepage}

\setcounter{page}{2} %please do not remove this line

%%%%%%%%%%%%%%%%%%%%%%%%%%%%%%%%
% begin main text
%%%%%%%%%%%%%%%%%%%%%%%%%%%%%%%%

\section{Introduction} 

Relativistic heavy--ion ({\mbox{A--A}\xspace}) collisions provide an excellent tool to study nuclear matter under extreme conditions of temperature and density and the phase transition between hadronic matter and a 
deconfined state of quarks and gluons, the quark--gluon plasma (QGP), predicted by lattice quantum chromodynamics (QCD) calculations~\cite{HotQCD:2014kol}. 
As the QGP produced in a heavy-ion collision expands, it cools down until a phase transition occurs that confines quarks and gluons inside hadrons around a temperature of 155--158~MeV~\cite{Andronic:2017pug}.
Soon after hadronisation, the resulting dense and hot gas of stable hadrons and resonances reaches the chemical freeze-out. Afterwards, hadrons keep interacting (pseudo-)elastically, and thus exchanging momentum, until the final decoupling at kinetic freeze-out where their momentum distribution is determined. In this stage, the hadron gas temperature decreases from around 150~MeV to approximately 100~MeV.
The late hadronic stage of heavy-ion collisions represents a unique environment for the study of a hadronic system in such conditions of high temperature and density. The understanding of the processes occurring in the excited hadron gas is therefore of primary relevance for the interpretation of observables that are employed to characterize the chemical and kinetic freeze-outs, the hadronic phase and more broadly speaking, the time evolution of a heavy-ion collision.
In this respect, short-lived resonances are sensitive probes to study the hadronic phase, which is formed after the chemical freeze-out. This is due to their different lifetimes that are comparable to the expected time duration of the hadronic phase ($\approx$ 1--10~fm/$c$)~\cite{Aamodt:2011mr}. 

The resonances with the shortest lifetime such as the {\ensuremath{{\rm \rho}^{0}}\xspace} (lifetime $c\tau \approx 1$ fm~\cite{Zyla:2020zbs}) decay during the hadronic phase and their decay products are subject to rescattering effects in the dense hadronic matter, losing memory of the resonance from which they have decayed. Due to the lack of correlation among the decay products, the result is that the resonance cannot be reconstructed via the usual invariant-mass analysis and the measured yield is suppressed with respect to expectations~\cite{Acharya:2018qnp}. The resonances with the longest lifetime such as the $\phi$ ($c\tau \approx$ 45~fm~\cite{Zyla:2020zbs}) are likely to survive the hadronic phase and decay in vacuum after the kinetic freeze-out. The scenario is further complicated by regeneration effects for which two particles in the hadronic phase interact via resonance formation, producing an enhancement with respect to the primordial production of that resonance.
The $\Sigma(1385)^{\pm}$ ($c\tau \approx$ 5.01~fm for $\Sigma(1385)^{-}$ and $c\tau \approx$ 5.48~fm for $\Sigma(1385)^{+}$~\cite{Zyla:2020zbs}) provides a crucial test case in this context. The typical modeling of the hadronic phase is based on transport codes in which hadronic cross sections are implemented to describe the known hadronic interactions. The widely used hadronic cascade simulator UrQMD~\cite{Bleicher:1999xi} is therefore coupled to event generators such as EPOS~\cite{Drescher:2000ha, Werner:2010aa} in order to predict the production of hadronic resonances. So far, this picture has been tested mostly with the {\ensuremath{{\rm K}^{*}(892)^{0}}\xspace} resonance which has a lifetime of $c\tau \approx$ 4.2~fm and its
production has been measured to be strongly suppressed~\cite{Acharya:2019qge} due to the dominance of rescattering effects. Results from {\ensuremath{{\rm \rho}^{0}}\xspace} and {\ensuremath{{\rm \Lambda} (1520)}\xspace} further support this picture of rescattering~\cite{Acharya:2018qnp, ALICE:2018ewo}. Despite its relatively short lifetime, the $\Sigma(1385)^{\pm}$ is expected to be essentially  unaffected by the hadronic phase, likely due to larger cross sections for the regeneration processes~\cite{Knospe:2016, Oliinychenko:2021enj}. 
In this Letter, the first measurement of $\Sigma(1385)^{\pm}$ in heavy-ion collisions at the LHC is reported, based on a data sample of \PbPb~collisions at a centre-of-mass energy per nucleon pair of $\sqrt{s_{\mathrm{NN}}}= 5.02$ TeV. The measurement presented in this Letter provides an essential test of our current understanding of the evolution of hadronisation induced by heavy-ion collisions.

This Letter is organised as follows. After a brief description of the ALICE detector and the data analysis in Section~\ref{sec.:Analysis}, the measured transverse-momentum (\pt) spectra and $p_{\rm T}$-integrated yields and ratios are presented in Section~\ref{sec.:Results} together with a discussion of our findings. Our conclusions are then summarized in Section~\ref{sec.:Conclusions}.

\section{Experiment and data analysis}
\label{sec.:Analysis}

\subsection{Experimental setup and event selection}
A detailed description of the ALICE detector and its performance can be found in Refs.~\cite{Aamodt:2008zz, ALICE:2014sbx}. This apparatus is optimized for providing particle identification (PID) in a wide momentum range (0.1--20 \GeVc) and high track-density environment by using different techniques. For this analysis, the Inner Tracking System (\ITS) and the Time Projection Chamber (\TPC) are used for vertex determination and tracking, while the TPC and the Time-Of-Flight (\TOF) systems are employed for PID.
These detectors cover the full azimuth over a pseudorapidity region \etarange{0.9} and are located inside a large solenoidal magnet providing a field of 0.5~T.
The \ITS~\cite{Aamodt:2010aa}, located at a radial distance $3.9~<~r~<~43$~cm from the beam axis, consists of two layers of silicon pixels (SPD), two layers of silicon drift chambers and two layers of silicon strips. The SPD, in particular, is used to reconstruct the track segments that serve to determine the primary vertex of the collision.
The \TPC~\cite{Alme:2010ke} is a large cylindrical drift chamber covering a radial distance $85~<~r~<~247$~cm and the main tracking device in the central barrel. 
The \TOF~\cite{Akindinov:2013tea} consists of a cylindrical array of MRPCs located at a radial distance of about 380~cm from the beam axis, with an intrinsic resolution of 50~ps.
Charged particles can be identified via their specific energy loss, \dEdx, measured in the TPC with a resolution of 5$\%$, and via their time-of-flight measured by TOF.

On either sides of the interaction point, two scintillator hodoscopes, the \VZEROA ($2.8~<~\eta~<~5.1$) and \VZEROC ($-3.7~<~\eta~<-1.7$)~\cite{Abbas:2013taa}, serve for triggering, background rejection and centrality classification, as shown in Table 3. 
In addition, two Zero Degree Calorimeters (\ZDC)~\cite{Arnaldi:2017dmc} placed at equal distances, 112.5~m, from the interaction region are used for background rejection and spectator nucleon measurements.  

The data samples analysed were recorded in 2018 using \PbPb~collisions at \mbox{$\sqrt{s_{\mathrm{NN}}}~=~5.02$}~TeV.
The minimum bias (MB) interaction trigger during the data collection requires a coincidence of signals in the \VZEROA and \VZEROC. The central and semi-central triggers require in addition to the MB trigger, an online selection on the total signal amplitude in the V0 detectors, corresponding to collision centralities of 0--10\% and 30--50\%, respectively. Background events are rejected using the timing information from the \ZDC and the \VZERO detectors~\cite{Abelev:2013qoq}. For the pile-up removal, the correlations between the number of reconstructed space points (clusters) in the \SPD, the number of reconstructed \SPD track segments and the total signal in the \VZEROA and \VZEROC detectors are considered. 

Events are selected only if they contain at least two tracks and a reconstructed primary vertex located within $\pm 10~$~cm with respect to the nominal interaction point along the beam axis ($z$ axis).
A total of $26\times10^{7}$ collision events have been accepted for the analysis.
Events are further classified into centrality classes, expressed in terms of percentiles of the total hadronic \PbPb cross section, using the signal amplitudes in the \VZERO arrays~\cite{Abelev:2013qoq, Adam:2015ptt, ALICE-PUBLIC-2018-011}. 
The analysis is performed in three centrality classes: 0--10\%, 30--50\% and 50--90\%. The 0--10\% and 50--90\% centrality classes correspond to the most central and the most peripheral \PbPb collisions with small and large impact parameters, respectively.

\subsection{Resonance reconstruction and yield extraction}
The resonances $\Sigma(1385)^{+}$ and $\Sigma(1385)^{-}$ are measured separately via their hadronic decay channel to \lmb and {\ensuremath{\pi^{\pm}}\xspace} with a branching ratio of 87\%~\cite{Zyla:2020zbs}, by calculating their invariant mass. The \lmb decays weakly into a p$\pi$ pair with a branching ratio of 63.9$\%$ and a characteristic \Vdecay topology, which is exploited for its reconstruction. In the following, when referring to $\Sigma(1385)^{+}$ or $\Sigma(1385)^{-}$, the sum of the particle and antiparticle (cc, charge conjugate) is considered.

Primary charged tracks are required to fulfill the standard criteria for good reconstruction quality described in Ref.~\cite{Acharya:2019uia}. These selections affect the {\ensuremath{\pi^{\pm}}\xspace} from the $\Sigma(1385)^{\pm}$ decays, for which it is additionally required to have $p_{\rm T} > 120$~\MeVc. Pion identification is based on the requirement that the \dEdx and the time-of-flight measured in the \TPC and in the \TOF, respectively, are compatible with their expected values within a fixed number of standard deviations ($\sigma_{~\rm TPC}$, $\sigma_{~\rm TOF}$)~\cite{Acharya:2019yoi}. In particular, the signals are requested to be within $\pm$5$\sigma_{~\rm TPC}$, $\pm$3$\sigma_{~\rm TPC}$ and $\pm$2$\sigma_{~\rm TPC}$ for tracks with $p_{\rm T} < 0.35$~\GeVc, $0.35 < p_{\rm T} < 0.5$~\GeVc, and $0.5 < p_{\rm T} < 20$~\GeVc, respectively.
If the measurement of the time-of-flight is available, this information is used in combination with the TPC PID by applying a selection based on a $\pm$3$\sigma_{~\rm TOF}$ range if $p_{\rm T} < 1.5$~\GeVc and on a $\pm$2.5$\sigma_{~\rm TOF}$ range otherwise.

For the decay products (daughters) of the \lmb, a subset of the standard track quality criteria are applied 
with the addition of $p_{\rm T} > 150$~\MeVc. 
Furthermore, the candidate {\lmb}s fulfill the criteria of the \Vdecay decay topology listed in Table~\ref{table:lambdacriteria}. The distance of closest approach (DCA) of the \lmb daughters is measured in standard deviations of this distribution. The pointing angle is the angle between the direction of the momentum of the \lmb and the line connecting the secondary to the primary vertex. The invariant mass window is defined so as to take into account the resolution of the reconstructed \lmb and $R_{\Lambda}$ is the allowed radial distance within which the \lmb decay is considered.
The \lmb daughters are identified as pions and protons based on a 4$\sigma_{~\rm TPC}$ criterium for all \pt.

\begin{table}[htp]
\centering
\caption{Selection criteria of the \lmb daughter of $\Sigma(1385)^{\pm}$.}\label{table:lambdacriteria}
    \begin{tabular}{ l  l }
    \hline
         $|y_{\lmb}|$ & $~<~0.5$ \\
         DCA of the proton daughter to primary vertex &  $>~0.12$~cm\\
         DCA of the pion daughter to primary vertex  & $>~0.3$~cm\\
         DCA between $\Lambda$ decay products & $<~0.8$ standard deviations\\
         $\lmb$ cosine of pointing angle & $>~$~0.98\\
         $\lmb$ invariant mass window & $m_{\rm PDG}~\pm 6.7$~\MeVmass\\
         $\lmb$ decay radius ($R_{\Lambda}$) & $5~<~R_{\Lambda}~<~200$~cm \\
         Proper lifetime & $<~25$~cm/$c$\\ 
         Daughters $p_{\rm T}$ & $>~150$ \MeVc \\ \hline
    \end{tabular}
\end{table}

The \lmb {\ensuremath{\pi}\xspace} pairs are reconstructed within the region \etarange{0.8} and \yrange{0.5} by combining candidates from the same event. The uncorrelated combinatorial background is estimated via the mixed-event technique, by combining \lmb and pions belonging to different events of the same centrality class with similar primary vertex position along the $z$-axis and charged-particle multiplicity. Specifically, the difference among the $z$-coordinate of the vertices of the events being mixed must be less than 1 cm and the difference of the charged-particle multiplicities less than five. To reduce the statistical fluctuations in the mixed-event background distribution, each event is mixed with nine other events.

As the $\Sigma(1385)^{\pm}$ decay products have a large mass difference, a selection on their momenta is applied which helps to reduce the combinatorial background under the peak. This is achieved by requiring the momentum asymmetry of the decay products (defined as ($p_{\lmb}-p_{\pi}$)/($p_{\lmb}+p_{\pi}$)) to be between 0.3 and 0.95. The asymmetry selection results in the partial removal of the correlated background remaining after the event-mixing background subtraction. 

The criteria used to select the \lmb {\ensuremath{\pi}\xspace} pairs are applied to both same-event and mixed-event invariant mass distributions, which are then normalized in the region 1.8$-$2.0~\GeVmass. The normalised mixed-event background distributions are subtracted from the same-event ones. The resulting invariant mass ($M_{\Lambda \pi}$) distributions exhibit the signal peak on top of a residual background consisting of correlated \lmb {\ensuremath{\pi}\xspace}  pairs from other particles or misidentified ones~\cite{Abelev:2014qqa}. The residual background has a smooth shape that is described by the following function:

\begin{equation}\label{eq:bck}
f_{\rm BG}=[M_{\Lambda \pi}-(m_{\pi}+m_{\Lambda})]^{n} ~ \mathrm{exp}(A+B \times M_{\Lambda \pi}+C \times M_{\Lambda \pi}^{2}),
\end{equation} 

where $m_{\pi}$ and $m_{\Lambda}$ are the $\pi$ and \lmb mass, respectively, taken from Ref.~\cite{Zyla:2020zbs} and $A$, $B$, $C$ and $n$ are free parameters.

\begin{figure}[tp]
\centering
\begin{minipage}[]{.5\textwidth}
  \centering
  \includegraphics[width=1.\linewidth]{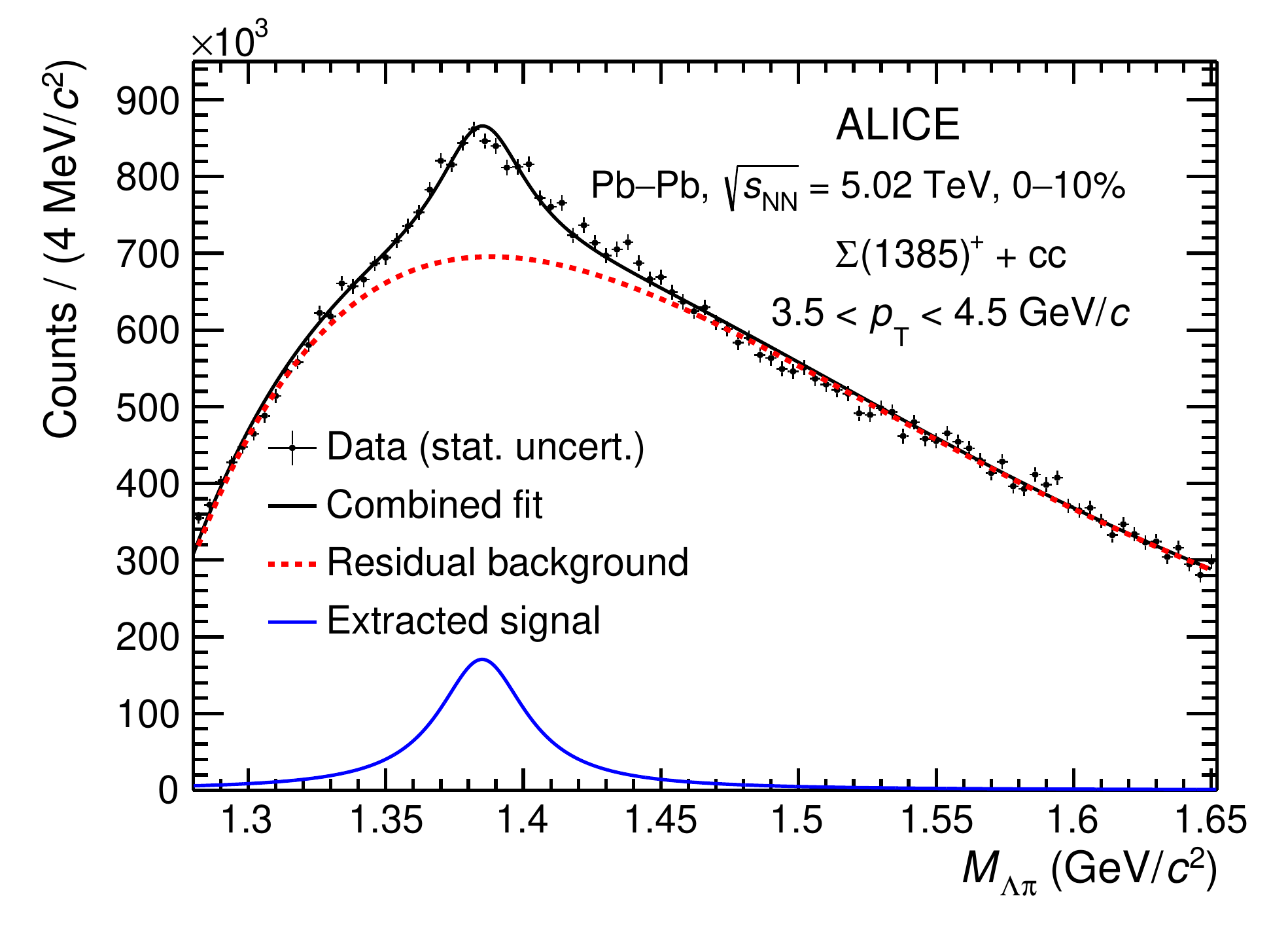}
\end{minipage}%
\begin{minipage}[]{.5\textwidth}
  \centering
  \includegraphics[width=1.\linewidth]{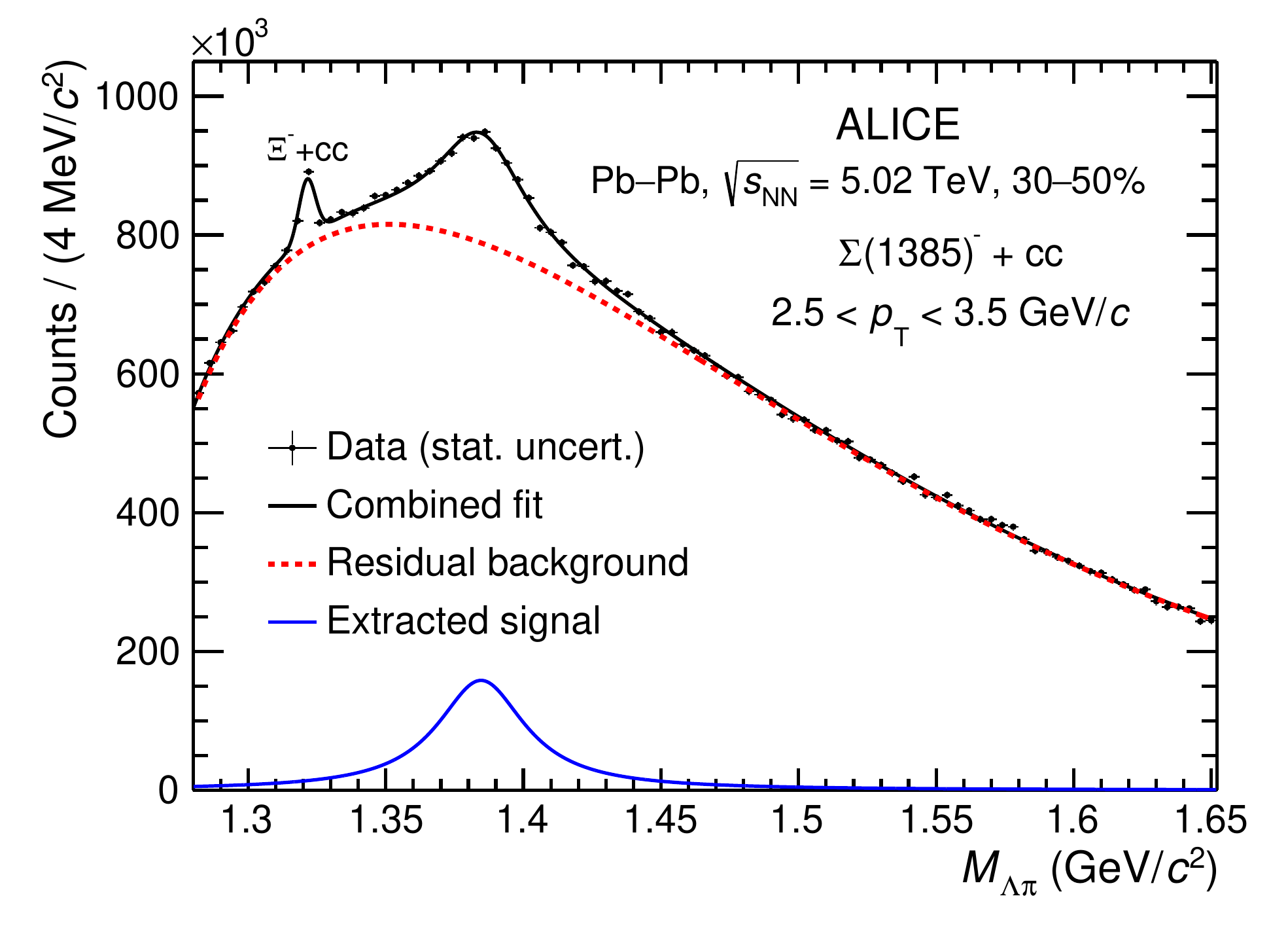}
\end{minipage}
\caption{Left:Invariant mass distribution of \lmb {\ensuremath{\pi}\xspace} pairs for $\Sigma(1385)^{+}$ after subtraction of the mixed-event background in 0--10$\%$ central \PbPb collisions and $3.5 < p_{\rm T} < 4.5$~\GeVc. \\Right:Invariant mass distribution of \lmb {\ensuremath{\pi}\xspace} pairs for $\Sigma(1385)^{-}$ after subtraction of the mixed-event background in 30--50$\%$ centrality class and $2.5 < p_{\rm T} < 3.5$~\GeVc. The black curves represent the combined fit using a signal (continuous blue line) plus residual background (red dashed line) model, as described in the text.} \label{fig:invMass}
\end{figure}

The mixed-event subtracted invariant mass distribution is fitted with a Breit--Wigner function for the signal and Eq.~\ref{eq:bck} for the residual background. The width of the resonances is kept fixed at the PDG value of 36~\MeVmass for $\Sigma(1385)^{+}$ and 39~\MeVmass for $\Sigma(1385)^{-}$~\cite{Zyla:2020zbs}. This procedure is repeated in nine intervals in the pair $p_{\rm T}$, from 1 to 9~\GeVc, and for each centrality class. The fit range for each $p_{\rm T}$ interval varies in order to achieve a better $\chi^{2}$ per degree of freedom. The lower values vary between 1.26 and 1.30~\GeVmass and the upper boundaries from 1.55 to 1.70~\GeVmass. The yields of $\Sigma(1385)^{\pm}$ are extracted in each $p_{\rm T}$ interval and centrality class by integrating the Breit--Wigner function in the range [$M_{\rm p}-5\Gamma, M_{\rm p}+5\Gamma$], where $M_{\rm p}$ is the peak position resulting from the fit and $\Gamma$ the width of the resonance.
In Fig.~\ref{fig:invMass}, the mixed-event subtracted invariant mass distribution is presented for $\Sigma(1385)^{+}$ (left) in the 0--10\% centrality class for $3.5 < p_{\rm T} < 4.5$~\GeVc and for $\Sigma(1385)^{-}$ (right) in the 30--50\% centrality class for $2.5 < p_{\rm T} < 3.5$~\GeVc. In the latter, the $\Xi$ peak is visible at $\approx$~1.321~\GeVmass, requiring for $\Sigma(1385)^{-}$ an additional gaussian function to be used in the fit to take into account the $\Xi$ particle. 

\subsection{Corrections}
\begin{figure}[htp]
\centering
\begin{minipage}[]{.5\textwidth}
  \centering
  \includegraphics[width=1.\linewidth]{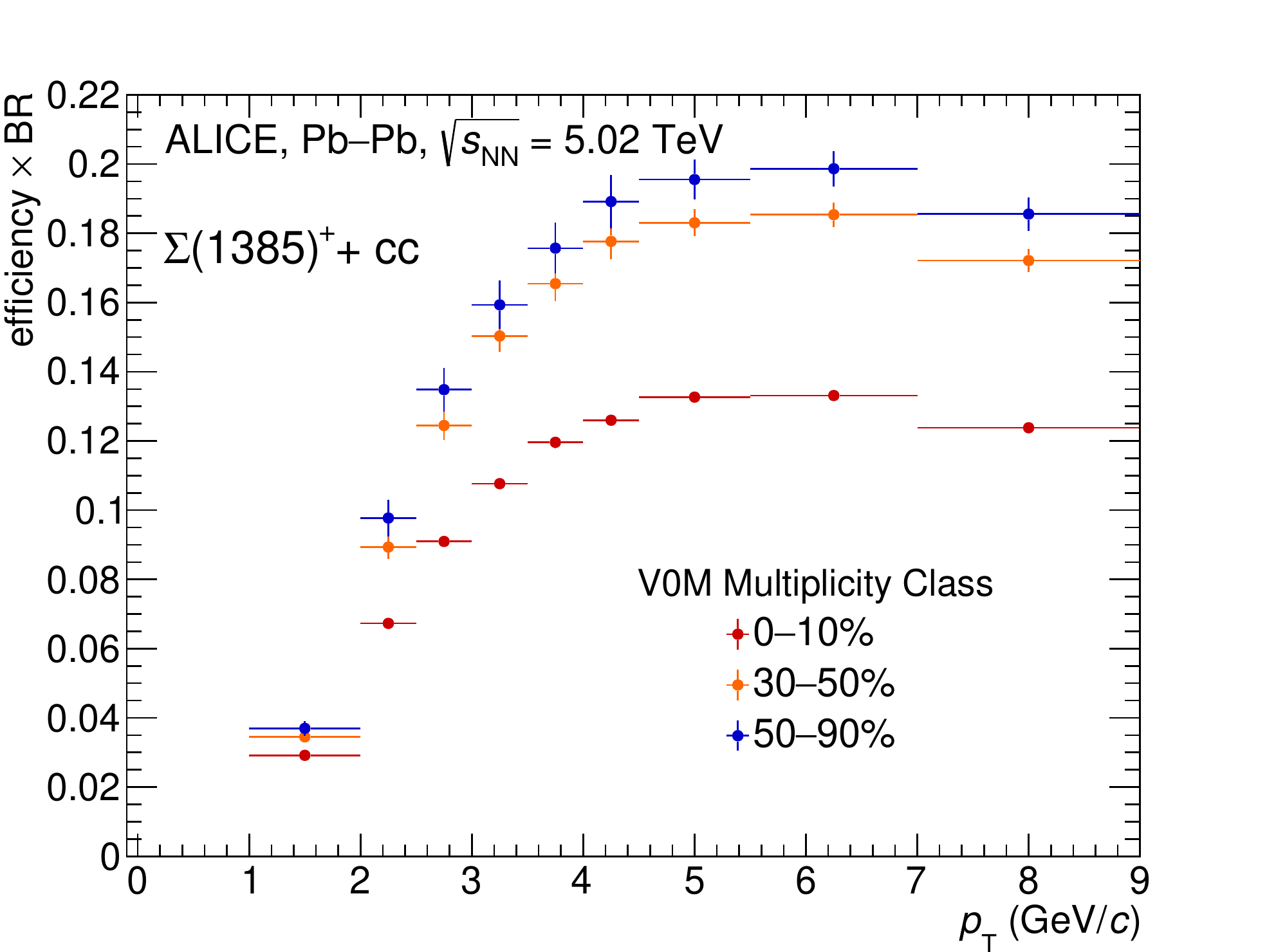}
\end{minipage}%
\begin{minipage}[]{.5\textwidth}
  \centering
  \includegraphics[width=1.\linewidth]{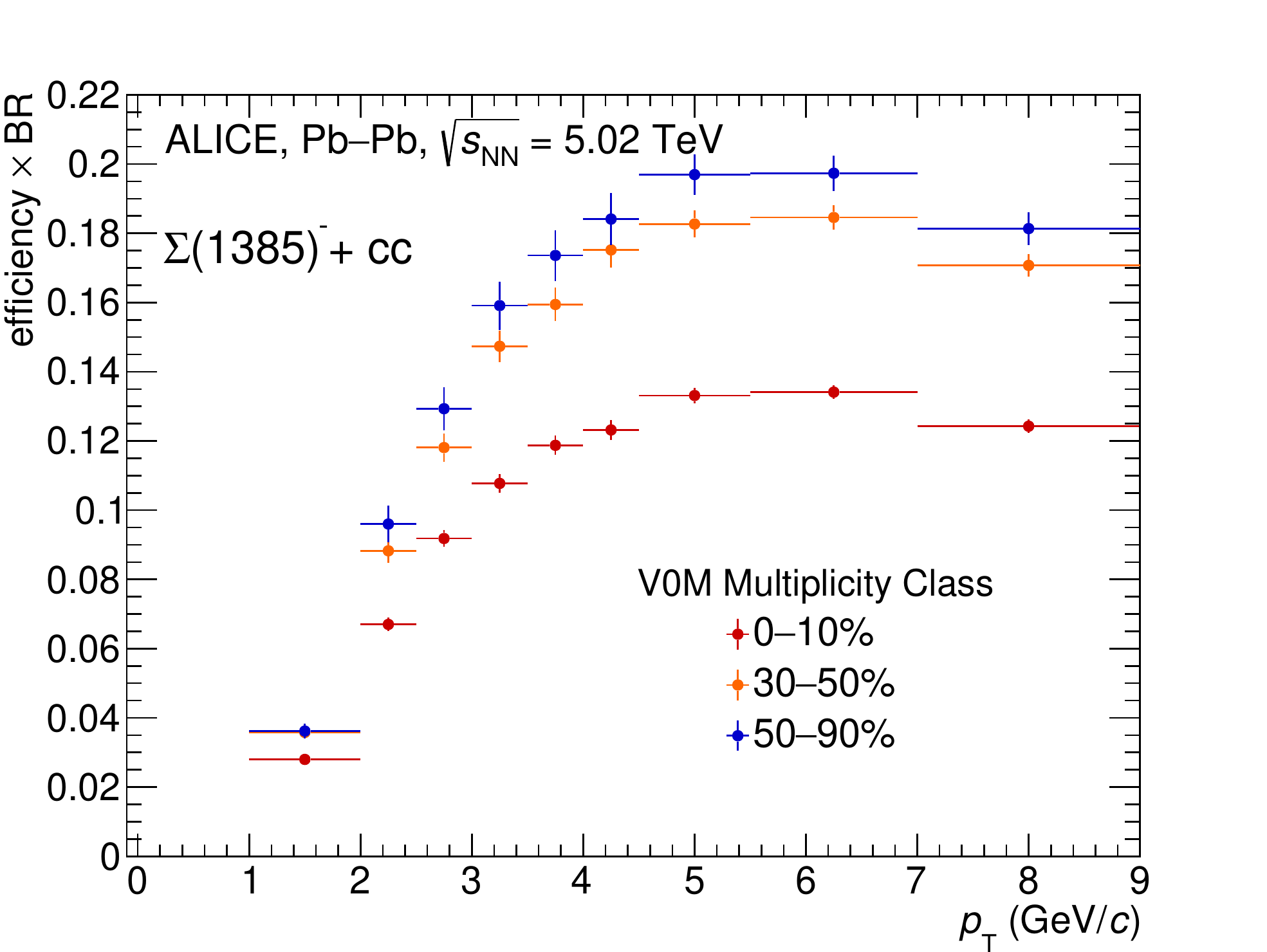}
\end{minipage}
\caption{The product of efficiency$\times$BR for $\Sigma(1385)^{+}$ (left) and $\Sigma(1385)^{-}$ (right) in the three centrality classes used for the analysis. The branching ratio, BR, is included in the correction. The error bars represent the statistical uncertainties.} \label{fig:corr}
\end{figure}

The extracted raw yields of $\Sigma(1385)^{\pm}$ are normalized to the number of events of the corresponding centrality class and corrected for  the detector acceptance ($A$), the reconstruction efficiency ($\epsilon_{\rm rec}$) and the branching ratio (BR)~\cite{Zyla:2020zbs}. 
The detector acceptance and reconstruction efficiency (called efficiency) are determined from a Monte Carlo simulation based on the HIJING event generator~\cite{hijing}. 
The $\Sigma(1385)^{\pm}$ signals are injected with a flat $p_{\rm T}$ distribution in the range 0$-$10~\GeVmass into HIJING events. 
The generated particles and their decay products are propagated through the detector material using GEANT3 as transport code~\cite{Geant:1994zzo}. 
The same criteria are applied to select the resonance decay products and the pairs as for the data analysis. The product $A\times\epsilon_{\rm rec} \times$BR is calculated as the fraction of the generated $\Sigma(1385)^{\pm}$ at midrapidity, reconstructed and identified after the application of all selection criteria. 
The calculation is performed in each centrality class and in Fig.~\ref{fig:corr}, the $A\times\epsilon_{\rm rec} \times$BR (denoted as efficiency$\times$BR) is shown as a function of $p_{\rm T}$ for the three centrality classes. It depends on the centrality due to the centrality dependence of the efficiency of both $\pi$ and \lmb particles. Due to the flat input $p_{\rm T}$ spectrum of the injected $\Sigma(1385)^{\pm}$ resonances, an iterative procedure is applied for re-weighting it, to remove input spectrum shape effects. 
The efficiencies obtained from the 
above described procedure are compared with those obtained from standard HIJING simulations and their ratio is compatible with unity.

\subsection{Systematic uncertainties}
The sources of systematic uncertainties in the measurement of the $\Sigma(1385)^{\pm}$ resonances are the yield extraction, the tracking efficiency of the pion daughter, the selection criteria of \lmb particles, the pion PID, the knowledge of the detector material budget, the asymmetry selection of the resonance daughter particles and the branching ratio to the decay channel used in the analysis. 
The uncertainties, except those of the asymmetry and the branching ratio, are $p_{\rm T}$ and centrality dependent. The uncertainty on the raw yield extraction comprises variations of the fit range, of the background fit function and of the mixed-event background normalization range, the use of bin counting instead of the integration of the signal function, as well as letting the width of the Breit--Wigner free in the fit. From all variations, the maximum deviation from the nominal value is assigned as systematic uncertainty. 
The difference between the global tracking efficiency in the data and the Monte Carlo simulation contributes to the total uncertainty by affecting the pion daughter and it varies from 5\% at \pt $=$ 1 GeV/$c$ to 3\% at \pt $=$ 6 GeV/$c$.  
The uncertainty due to the \lmb topological selection is calculated by varying the DCA of the \lmb daughters to the primary vertex, the DCA of the \lmb daughters and the cosine of pointing angle. This uncertainty has a value ranging from 2--3\% at the lowest $p_{\rm T}$ to 4--5\% at the highest $p_{\rm T}$. The uncertainty associated with the pion identification is quantified by varying the selections in the TPC and the TOF detectors, and  reaches a maximum value of 8\% for $\Sigma(1385)^{-}$ in central collisions. The uncertainty on the yields occurring from the implementation of the material budget in the detector simulation was evaluated by increasing and decreasing the material amount within its systematic uncertainty~\cite{ALICE:2014sbx}, resulting in a variation of $\pm 4.5$\%. Reconstruction in data and Monte Carlo was done in the two extreme cases and the systematic uncertainty has been estimated on the final observables, which are the \pt distributions of the resonances. The asymmetry uncertainty is 5\%, regardless of $p_{\rm T}$ and centrality, and is evaluated by varying the accepted asymmetry range. Finally, the branching ratio uncertainty is 1.1\%~\cite{ALICE:2017pgw}. In Table~\ref{table:resSys}, the uncertainties entering the measurement are given for two $p_{\rm T}$ intervals in 0--10$\%$ central collisions. The total systematic uncertainty of $\Sigma(1385)^{-}$ is slightly higher than the uncertainty of $\Sigma(1385)^{+}$ mainly due to the fact that in the yield extraction of the former an additional gaussian function has to be taken into account in the fit to parametrise the $\Xi$ peak. In the other centrality classes the uncertainties are lower.

\begin{table}[htp]
\caption{Sources of systematic uncertainties on the $\Sigma(1385)^{\pm}$ resonance yields in 0--10$\%$ most central \PbPb collisions at \fivenn, given for two different \pt intervals, 1--2 \GeVc and 5.5--7 \GeVc.}\label{table:resSys}
\begin{center}
    \begin{tabular}{ | c | c | c |}
    \hline
         Systematic variation & $\Sigma(1385)^{+}$ + cc & $\Sigma(1385)^{-}$ + cc \\ \hline \hline
          & $p_{\rm T}$ (\GeVc)  & $p_{\rm T}$ (\GeVc) \\ \cline{2-3}
                              &      1--2~~~~5.5--7 &  1--2 ~~~~ ~5.5--7     \\ \hline 
         Yield extraction (\%)          & 13.0~~~~13.2 &  15.5 ~~~~~~13.6    \\ 
         Global tracking efficiency (\%) & ~~5.0 ~~~~~3.0 & ~~5.0 ~~~~~~~~3.0       \\ 
         \lmb topological cuts (\%)   & ~~3.3 ~~~~~5.1     &  ~~2.6 ~~~~~~~~4.3      \\ 
         Pion PID (\%)   & ~~4.1 ~~~~~4.2     &  ~~2.1 ~~~~~~~~7.7      \\ 
         Material budget (\%)    & ~~5.2 ~~~~~1.5     &  ~~5.2 ~~~~~~~~1.5      \\ 
         Daughters asymmetry (\%)    & ~~5.0 ~~~~~5.0     &  ~~5.0 ~~~~~~~~5.0      \\ 
         Branching ratio (\%)    & ~~1.1 ~~~~~1.1     &  ~~1.1 ~~~~~~~~1.1      \\ \hline
         Total (\%)              & ~16.0 ~~~15.8     & ~17.6 ~~~~~~17.1       \\ \hline 
         Total multipl. uncorrelated (\%)  & ~~~9.8 ~~~~~5.3     & ~~9.7 ~~~~~~~~~6.1       \\ \hline
         
    \end{tabular}
\end{center}
\end{table}

\section{Results and Discussion}
\label{sec.:Results}
\begin{figure}
\centering
\begin{minipage}[]{.5\textwidth}
  \centering
  \includegraphics[width=1.09\linewidth]{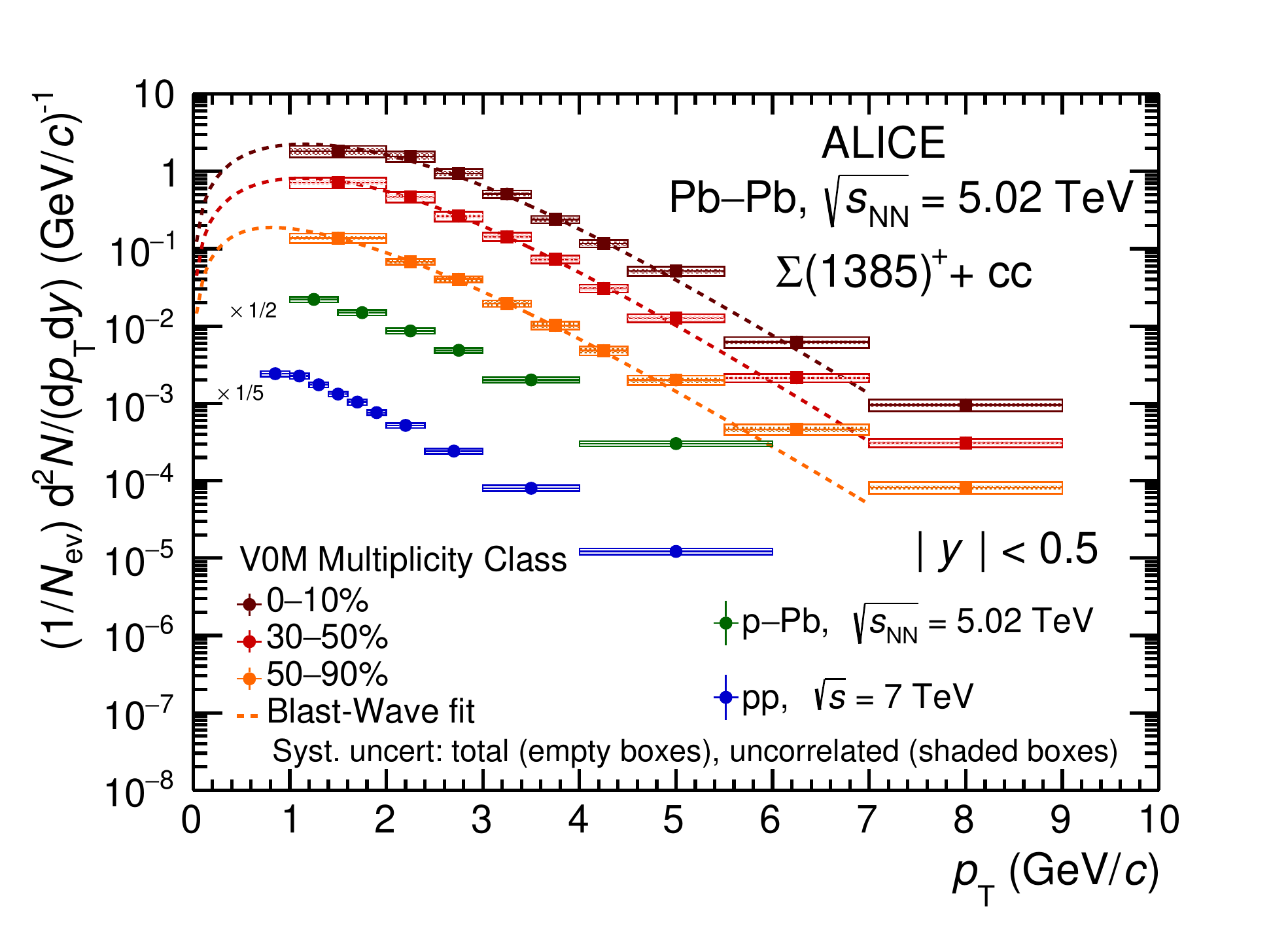}
\end{minipage}%
\begin{minipage}[]{.5\textwidth}
  \centering
  \includegraphics[width=1.09\linewidth]{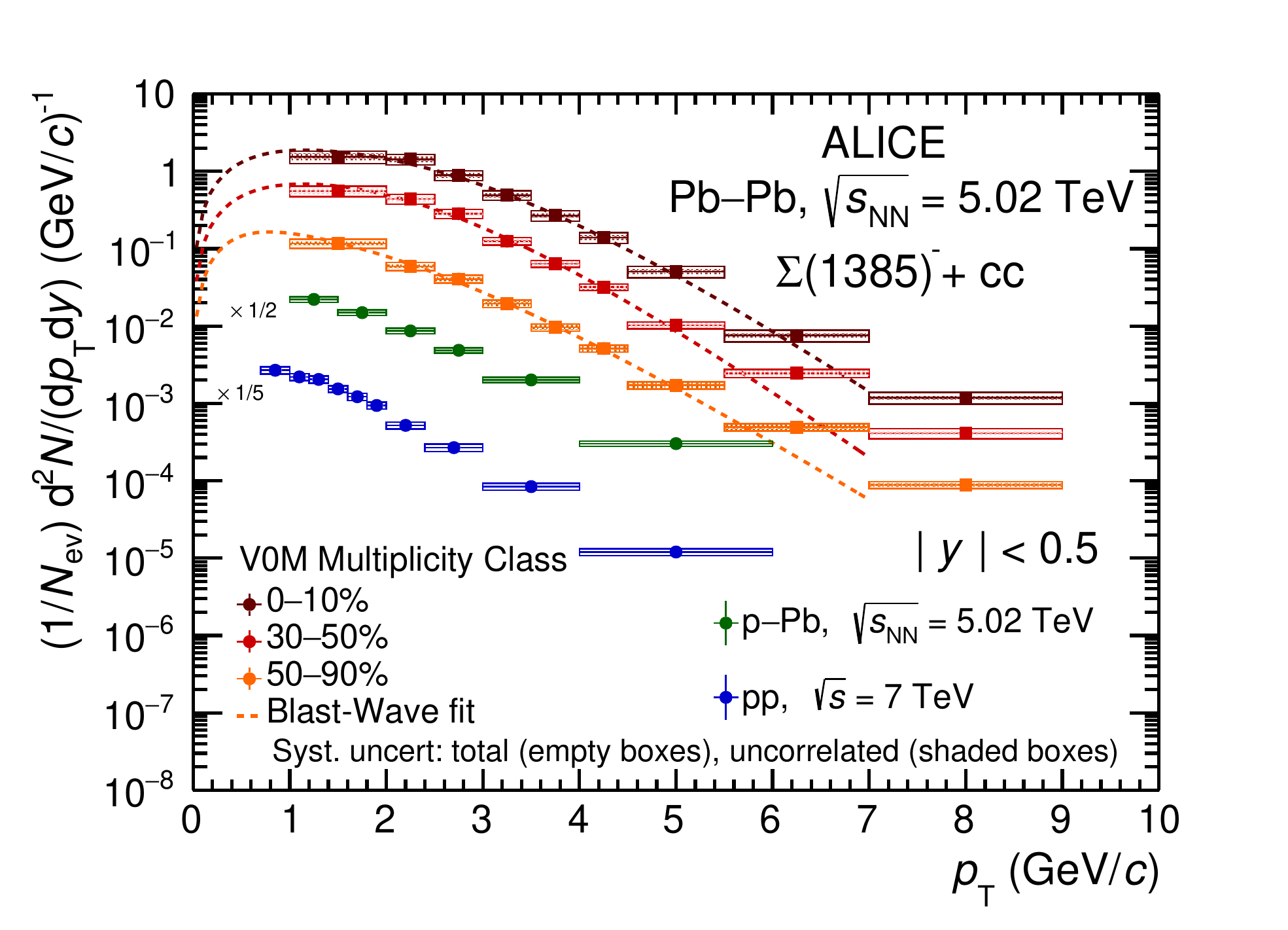}
\end{minipage}
\caption{$p_{\rm T}$ spectra of $\Sigma(1385)^{+}$ (left) and $\Sigma(1385)^{-}$ (right) resonances in \PbPb collisions at \fivenn in three centrality classes. Empty and shaded boxes depict the total and uncorrelated uncertainties, respectively. The Blast-Wave fit functions are plotted up to 7 \GeVc. The measurements in \pp collisions at \seven~\cite{Abelev:2014qqa} and in \pPb collisions at \five (NSD)~\cite{ALICE:2017pgw} are quoted for comparisons.}\label{fig:sigmaSpectra}
\end{figure}

The transverse-momentum distributions of $\Sigma(1385)^{+}$ and $\Sigma(1385)^{-}$ in \PbPb~collisions are reported in Fig.~\ref{fig:sigmaSpectra} and compared to the measurements in inelastic \pp collisions at \seven~\cite{Abelev:2014qqa} and in non-single diffractive (NSD) \pPb collisions at \fivenn~\cite{ALICE:2017pgw}. 

Each of the \PbPb~distributions in this figure is fitted individually with a Blast-Wave function~\cite{Schnedermann:1993ws}. The $p_{\rm T}$-integrated yields and the mean transverse momentum, \meanpt, are calculated based on the data where the spectra are measured and on the extrapolation of the Blast-Wave function at low $p_{\rm T}$. The fraction of the integrated yields in the high-$p_{\rm T}$ extrapolation region is negligible, while this fraction for the low-$p_{\rm T}$ one ranges from 28\% to 37\%, depending on the collision centrality. The statistical and systematic uncertainties of the $p_{\rm T}$-integrated yields and the mean $p_{\rm T}$ are evaluated by repeating the fit after moving the spectra within their statistical and systematic uncertainties, respectively. The procedure of the fitting and the calculation of the $p_{\rm T}$-integrated yields and the mean $p_{\rm T}$ described above is repeated with Boltzmann, Fermi-Dirac, $m_{\rm T}$-exponential and \LT~\cite{Tsallis:1987eu} functions to calculate an additional systematic uncertainty defined as the maximum deviation of these quantities from the ones obtained with the Blast-Wave fit.  
The d$N$/d$y$ and \meanpt are reported in Tables~\ref{table:intYields} and~\ref{table:meanPt}, respectively, for the three centrality classes of this analysis. As a part of the total systematic uncertainty is correlated across multiplicity, the uncorrelated part of the uncertainties has been calculated by following a similar procedure as the calculation of the total uncertainty. The multiplicity-uncorrelated uncertainties are represented by the third number in Tables 3 and 4. The mean charged-particle multiplicity density for each centrality class, \avdndeta, is also reported in Table~\ref{table:intYields}. Both d$N$/d$y$ and \meanpt increase with \avdndeta in \PbPb collisions.

From \pp~to central heavy-ion collisions, the spectra become harder with increasing charged-particle multiplicity, following a similar trend as observed for the other measured light-flavour particles and resonances~\cite{Acharya:2019yoi, Acharya:2019qge}. The \meanpt increases from about 1.15 \GeVc in \pp collisions to about 1.37 \GeVc in \pPb and further increases with centrality in \PbPb as reported in Table 4.
\begingroup
\setlength{\tabcolsep}{1pt}
\begin{table}[htbp]
\centering
\caption{Average charged-particle multiplicity density per centrality class at midrapidity ($\etarange{0.5}$) and \\$p_{\rm T}$-integrated yields for $\Sigma(1385)^{\pm}$. The first, second and third  uncertainty in the $p_{\rm T}$-integrated yields indicate the statistical, the total systematic uncertainty and the multiplicity-uncorrelated systematic uncertainty, respectively.}\label{table:intYields}
%\begin{left}
 
    \begin{tabular}{c c c c}
    \hline
         Class & $\langle \dndeta\rangle$ & \multicolumn{2}{c}{d$N$/d$y$} \\ \cline{3-4}
          &  &$\Sigma(1385)^{+}$ + cc&$\Sigma(1385)^{-}$ + cc \\ \hline 
         0--10\%& 1756.6$\pm$51.5 & 4.956$\pm$0.005$\pm$0.839$\pm$0.550 & 4.746$\pm$0.005$\pm$0.871$\pm$0.536  \\
         30--50\%& ~~415.0$\pm$13.5& 1.851$\pm$0.003$\pm$0.311$\pm$0.158& 1.599$\pm$0.002$\pm$0.266$\pm$0.160 \\
   50--90\% &~~~~85.4$\pm$ ~9.5& $(3.357\pm0.007\pm0.494\pm0.308)\times10^{-1}$ & $(2.800\pm0.006\pm0.471\pm0.251)\times10^{-1}$\\ \hline
    \end{tabular}
%\end{left}
\end{table}
\endgroup

\begin{table}[htpb]
\centering
\caption{\meanpt of $\Sigma(1385)^{\pm}$ per centrality class. The first, second, and third uncertainty indicate the statistical, the total systematic uncertainty and the multiplicity-uncorrelated systematic uncertainty, respectively. }\label{table:meanPt}
%\begin{left}
    \begin{tabular}{  c  c  c }
    \hline
         Class & \multicolumn{2}{c}{\meanpt (\GeVc)} \\ \cline{2-3}
          & $\Sigma(1385)^{+}$ + cc&$\Sigma(1385)^{-}$ + cc \\ \hline 
         0--10\% & $(17.185\pm0.007\pm0.812\pm0.748)\times 10^{-1}$& $(16.974\pm0.007\pm1.067\pm0.974)\times 10^{-1}$ \\
         30--50\%& 1.556$\pm$0.001$\pm$0.068$\pm$0.057& 1.587$\pm$0.001$\pm$0.082$\pm$0.075 \\
         50--90\% & 1.465$\pm$0.001$\pm$0.086$\pm$0.079 &1.551$\pm$0.002$\pm$0.103$\pm$0.070 \\ \hline
    \end{tabular}
%\end{left}
\end{table}

\begin{figure}[htpb]
\begin{center}
\includegraphics[scale=0.5]{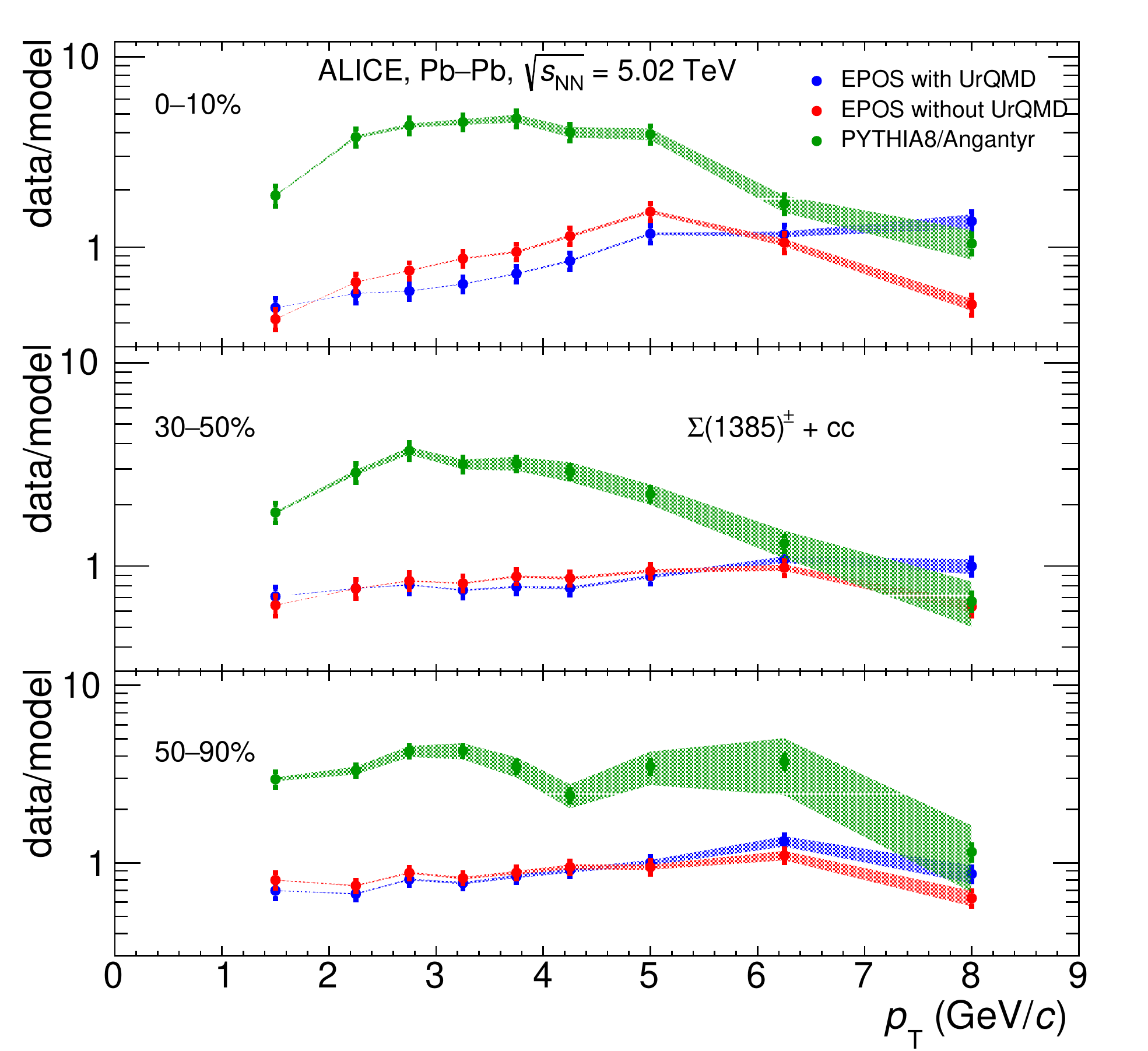}
\caption{Ratio of the measured \pt distributions of summed $\Sigma(1385)^{\pm}$ to model predictions from \Angantyr (green) and EPOS with (blue) and without (red) the UrQMD afterburner. Each panel corresponds to one of the three centrality classes in \PbPb collisions at \fivenn. Shaded bands represent the model statistical uncertainty, while bars correspond to the data total uncertainties.}\label{fig:PanelsRatioToEposAndBW}
\end{center}
\end{figure}

Figure~\ref{fig:PanelsRatioToEposAndBW} presents the ratios of the measured spectra of the 
summed $\Sigma(1385)^{\pm}$ resonance states to the
distributions obtained from the EPOS3 and \Angantyr Monte Carlo event generators in the three \PbPb centrality classes.
The EPOS3 model~\cite{Drescher:2000ha, Werner:2010aa, Werner:2013tya} describes the evolution of a heavy-ion collision with the reaction volume being divided into a core and a corona part. For high string densities, the model does not allow the strings to decay independently, instead, if the energy density from string segments is high enough, these fuse into the so-called ``core'' region, which evolves hydrodynamically. The low energy density region forms the ``corona'', which hadronizes according to the unmodified string fragmentation. After hadronization, hadrons are fed into the UrQMD hadron cascade afterburner, which describes hadronic interactions in a microscopic approach~\cite{Bleicher:1999xi}. 
Previous ALICE measurements of {\ensuremath{{\rm \rho}(770)^{0}}\xspace}, {\ensuremath{{\rm K}^{*} (892)^{0}}\xspace} and {\ensuremath{{\rm \Lambda} (1520)}\xspace} in \PbPb collisions at \twosevensixnn were compared with predictions from the EPOS 3.107 event generator~\cite{Knospe:2016, Acharya:2018qnp, ALICE:2018ewo}, indicating that the UrQMD afterburner is necessary for the model to describe the spectral shape of these resonances in central collisions, especially at low \pt.
Remarkably, the model calculation with UrQMD in~\cite{Knospe:2016} predicted that regeneration effects could balance rescattering in the hadronic phase for the $\Sigma(1385)^{\pm}$ decay products, resulting in no suppression for this resonance as a function of centrality.   
The new $\Sigma(1385)^{\pm}$ data reported in this Letter are compared with the latest version of EPOS3 (EPOS 3.4) both with and without coupling it to the UrQMD afterburner in Fig.~\ref{fig:PanelsRatioToEposAndBW}.
No significant difference is observed between the calculation with the UrQMD afterburner and without it in semicentral and  peripheral \PbPb collisions at \fivenn. A difference in the two predictions is observed for most central collisions for $1~<~\pt~<~5$~GeV/$c$. 
The model largely overestimates the production of $\Sigma(1385)^{\pm}$ resonances for $p_{\rm T} <$~5 GeV/$c$ in 0--10$\%$ central \PbPb collisions, $\approx 60\%$ in the lowest \pt~interval, whereas it describes the data within 20--30\% in semicentral and peripheral collisions.

\Angantyr~\cite{Bierlich:2018xfw} is an extension of the \Pythia8~\cite{Sjostrand:2014zea} event generator to the case of heavy-ion collisions. \Pythia describes nucleon--nucleon interaction at the parton level, based on multiple partonic interactions and Lund-string hadronization. In Angantyr, \Pythia is extended to model nucleon--nucleus and nucleus--nucleus collisions in a three steps procedure. First, the number of participating nucleons in a collision is calculated from a Glauber-like model. Second, \Pythia is used to simulate the collision at the partonic level according to the results of the Glauber calculation as a sum of incoherent inelastic, diffractive and/or elastic collisions. 
Third, the partonic state is allowed to have final state interactions before it is hadronized according to the Lund-string model in \Pythia. 
Hadronic rescattering and regeneration processes like those implemented in UrQMD are not considered in the version of \Angantyr employed here. 
The data-to-model ratios reported in Fig.~\ref{fig:PanelsRatioToEposAndBW} show that \Angantyr underpredicts the $\Sigma(1385)^{\pm}$ production at low momenta by a factor of 3 to 4 and exhibits a softer spectrum. It only tends to describe the data reasonably well at momenta above $p_{\rm T} \simeq 7$~GeV/$c$.

\begin{figure}[htp]
\begin{center}
\includegraphics[scale=0.5]{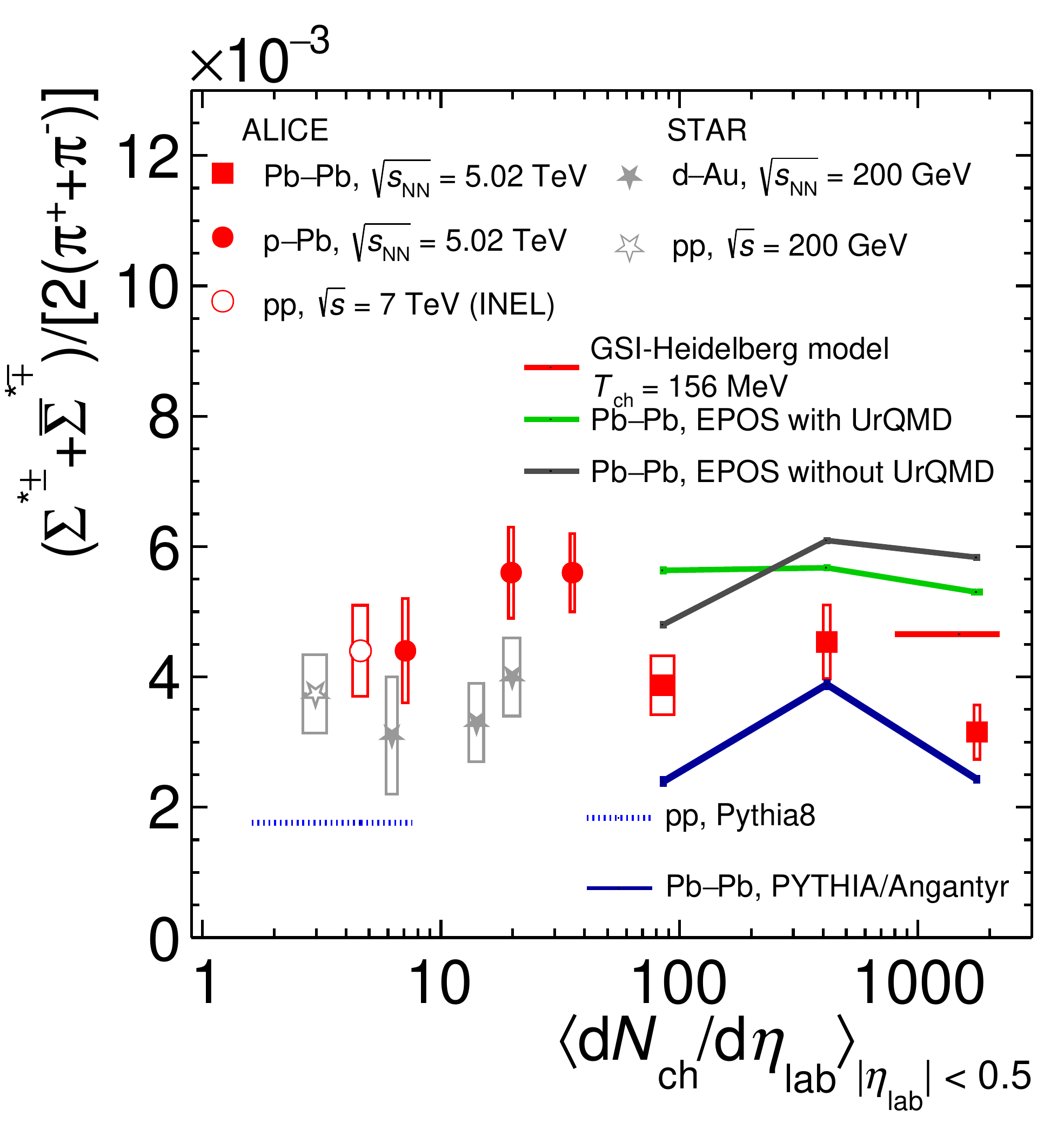}
\caption{$\Sigma(1385)^{\pm}$ to pion yield ratio measured in ALICE~\cite{Abelev:2014qqa, ALICE:2017pgw} together with the STAR measurements~\cite{Adams:2006br, Adams:2008ar} in various collision systems and energies are reported as a function of the charged-particle multiplicity density at midrapidity, ${\ensuremath{\langle{\ensuremath{\mathrm{d}N_\mathrm{ch}/\mathrm{d}\eta_{\rm lab}}\xspace}\rangle}_{|\mathrm{\eta_{\rm lab}<0.5}|}\xspace}$. Data are compared with model calculations for LHC energies from the GSI--Heidelberg grand canonical statistical hadronization model~\cite{Andronic:2017pug}, \Pythia8.2~\cite{Sjostrand:2014zea}, \Angantyr~\cite{Bierlich:2018xfw} and EPOS3 with and without UrQMD~\cite{Werner:2013tya}.} \label{fig:sigmaToPi}
\end{center}
\end{figure}

The $p_{\rm T}$-integrated yield ratios of $\Sigma(1385)^{\pm}$ to pions are shown in Fig.~\ref{fig:sigmaToPi} for different collision systems as measured by the ALICE~\cite{Abelev:2014qqa, ALICE:2017pgw} and the STAR collaborations~\cite{Adams:2006br, Adams:2008ar}. 
In general, no particular trend with multiplicity is observed given the uncertainties. A fit with a zeroth order polynomial of all data points reported in Fig.~\ref{fig:sigmaToPi} yields a $\chi^{2}/$NDF of 20.2/10, whereas the exclusion of the \PbPb most central point from the fit leads to a $\chi^{2}/$NDF of 14.6/9. In addition, for the ratio of the most central \PbPb point to the \pp one, we obtain $0.86\pm 0.16$ by taking into account both the statistical and systematic uncertainties, with the latter having the largest contribution to the error of this ratio.

In the same figure, models applicable to different charged-particle multiplicities are also presented. The comparison to the models, concerning especially the higher multiplicities, can give insight to the dominating mechanism in the hadronic phase that results in the observed behaviour. 
The statistical hadronization model~\cite{Andronic:2017pug} (indicated as GSI-Heidelberg in Fig.~\ref{fig:sigmaToPi}) describes the process of hadron formation at the scale where perturbative QCD is no longer applicable. It is assumed that near hadronization the fireball created in heavy-ion collisions is close to thermal equilibrium and hadron yields can be characterized by a grand canonical partition function. The prediction compared to data in Fig.~\ref{fig:sigmaToPi} is obtained for a chemical freeze-out temperature of T$=$156 MeV, which results from a fit to the light-flavour hadron and nucleus yields measured by ALICE at the LHC~\cite{Andronic:2017pug}.
For the $\rho(770)^{0}$~\cite{Acharya:2018qnp}, {\ensuremath{{\rm K}^{*} (892)^{0}}\xspace}~\cite{Acharya:2019qge} and {\ensuremath{{\rm \Lambda} (1520)}\xspace}~\cite{ALICE:2018ewo} short-lived resonances, a suppression with respect to the grand canonical statistical hadronization model expectation is observed for central \PbPb collisions, while the data in peripheral collisions are well described. This behaviour, which is also observed for $\Sigma(1385)^{\pm}$ in Fig.~\ref{fig:sigmaToPi}, is typically attributed to rescattering in the hadronic phase. 
In 0--10\% central \PbPb collisions, the $\Sigma(1385)^{\pm}$ to pion ratio is $3.6\sigma$ lower than the statistical hadronization model prediction, whereas the difference of the zeroth order polynomial fit (without taking into account the most central event class) from the same model is a $3.3\sigma$ effect.
The calculations from EPOS3 with UrQMD, which is able to describe qualitatively the suppression of the ${\rm K}^{*}(892)^{0}$/K and $\Lambda$(1520)/$\Lambda$ yield ratios from central to peripheral \PbPb collisions~\cite{Acharya:2018qnp, Acharya:2019qge, ALICE:2018ewo}, reproduce qualitatively also the weak centrality dependence of the $\Sigma(1385)^{\pm}/\pi$ yield ratio. Its magnitude, however, is overpredicted  for all centralities. 

It is to be noted that the \pt-integrated $\Sigma(1385)^{\pm}$/$\pi$ yield ratio (see Fig.~\ref{fig:sigmaToPi}) in the model exhibits a discrepancy with respect to data that is consistent with the one observed for the $\Sigma(1385)^{\pm}$ spectrum alone. In both cases, EPOS3 with UrQMD overestimates the production.

The centrality dependence of the $\Sigma(1385)^{\pm}/\pi$ ratio is qualitatively captured also by \Angantyr, within the current uncertainties.
As \Angantyr underpredicts the production at low transverse momenta of both $\Sigma(1385)^{\pm}$ and pions, it consequently underpredicts the $p_{\rm T}$-integrated production. Interestingly, this discrepancy is already present in pp collisions~\cite{Abelev:2014qqa}, indicating that it is not related to the heavy-ion modeling part. The semicentral \PbPb point is closer to \Angantyr prediction because the pion yield is underestimated to a greater degree than the sigma yield.

\section{Conclusions}
\label{sec.:Conclusions}

We have presented the first measurement of $\Sigma(1385)^{\pm}$ production in heavy-ion collisions at LHC energies. An evidence of suppression with respect to the grand canonical thermal model is observed in central collisions as for the {\ensuremath{{\rm K}^{*}(892)^{0}}\xspace} meson, which has a similar lifetime. 
While the EPOS3 model coupled to the UrQMD afterburner describes the centrality dependence of the {\ensuremath{{\rm K}^{*}(892)^{0}}\xspace} data, it clearly overestimates the production of $\Sigma(1385)^{\pm}$. This may either be caused by a missing element in the model that is common to all centralities or 
due to the fact that the rescattering and (the seemingly dominant) regeneration effects that EPOS3 withUrQMD predicts for $\Sigma(1385)^{\pm}$ do not manifest themselves in the data.
The current implementation of \Angantyr does not reproduce the $\Sigma(1385)^{\pm}$ yield either, while capturing the centrality dependence of $\Sigma(1385)^{\pm}/\pi$ within the uncertainties. It remains to be seen if ongoing developments of the model to include hadron rescattering will provide a better agreement with the data.

On the experimental side, future higher precision measurements will clarify if a suppression with respect to pp or peripheral \PbPb collisions is present and thus will allow for a model-independent investigation of the rescattering and regeneration picture. In addition, detailed comparisons with future $\Lambda$ measurements will elucidate if the observed data-model discrepancies are driven by the strangeness content of the hadron under study or the modeling of the hadronic phase.

%%%%%%%%%%%%%%%%%%%%%%%%%%%%%%%%
% end main text 
%%%%%%%%%%%%%%%%%%%%%%%%%%%%%%%%

%%%%% acknowledgements - handled by EB chairs 
\newenvironment{acknowledgement}{\relax}{\relax}
\begin{acknowledgement}
\section*{Acknowledgements}
% add specific acknowledgements here 
% ...but please don't remove the line below: funding agencies
% will be acknowledged with a custom tex file handled by EB chairs after Collab Round 2
\input{fa_2022-05-09_Opt_C.tex}
\end{acknowledgement}

%%%%%%%% Bibliography 
\bibliographystyle{utphys}   % Remember we use title in the biblio
\bibliography{main.bib}
%\input {bibliography.tex}  

%%%%%%%%%%%%%%%%%%%%%%%%%%%%%%%%
% Appendices: yours (if any) + authorlist
%%%%%%%%%%%%%%%%%%%%%%%%%%%%%%%%
\newpage
\appendix

%
%\input{} % put your appendices here (if any)
%

%%%%% Authorlist - please do not touch: handled by EB chairs 
\section{The ALICE Collaboration}
\label{app:collab}
\input{2022-05-09-Alice_Authorlist_2022-05-09_0_Opt_C.tex}  
\end{document}

%% file: commands.tex
%%%%%%%%%%%%%%%%%%%%%%%%%%%%%%%%%%%%%%%%%%%%%%%%%%
% These are some new commands that may be useful 
% for paper writing in general. If other newcommands
% are needed for your specific paper, please feel 
% free to add here. 
%
% The currently available commands are organized in: 
% 1) Systems
% 2) Quantities
% 3) Energies and units
% 4) Detectors
% 5) particle species 
%%%%%%%%%%%%%%%%%%%%%%%%%%%%%%%%%%%%%%%%%%%%%%%%%%

% 1) SYSTEMS 
\newcommand{\pp}           {pp\xspace}
\newcommand{\ppbar}        {\mbox{$\mathrm {p\overline{p}}$}\xspace}
\newcommand{\XeXe}         {\mbox{Xe--Xe}\xspace}
\newcommand{\PbPb}         {\mbox{Pb--Pb}\xspace}
\newcommand{\pA}           {\mbox{pA}\xspace}
\newcommand{\pPb}          {\mbox{p--Pb}\xspace}
\newcommand{\AuAu}         {\mbox{Au--Au}\xspace}
\newcommand{\dAu}          {\mbox{d--Au}\xspace}

% 2) QUANTITIES 
\newcommand{\s}            {\ensuremath{\sqrt{s}}\xspace}
\newcommand{\snn}          {\ensuremath{\sqrt{s_{\mathrm{NN}}}}\xspace}
\newcommand{\pt}           {\ensuremath{p_{\rm T}}\xspace}
\newcommand{\meanpt}       {$\langle p_{\mathrm{T}}\rangle$\xspace}
\newcommand{\ycms}         {\ensuremath{y_{\rm CMS}}\xspace}
\newcommand{\ylab}         {\ensuremath{y_{\rm lab}}\xspace}
\newcommand{\etarange}[1]  {\mbox{$\left | \eta \right |~<~#1$}}
\newcommand{\yrange}[1]    {\mbox{$\left | y \right |~<~#1$}}
\newcommand{\dndy}         {\ensuremath{\mathrm{d}N_\mathrm{ch}/\mathrm{d}y}\xspace}
\newcommand{\dndeta}       {\ensuremath{\mathrm{d}N_\mathrm{ch}/\mathrm{d}\eta}\xspace}
\newcommand{\avdndeta}     {\ensuremath{\langle\dndeta\rangle}\xspace}
\newcommand{\dNdy}         {\ensuremath{\mathrm{d}N_\mathrm{ch}/\mathrm{d}y}\xspace}
\newcommand{\Npart}        {\ensuremath{N_\mathrm{part}}\xspace}
\newcommand{\Ncoll}        {\ensuremath{N_\mathrm{coll}}\xspace}
\newcommand{\dEdx}         {\ensuremath{\textrm{d}E/\textrm{d}x}\xspace}
\newcommand{\RpPb}         {\ensuremath{R_{\rm pPb}}\xspace}

% 3) ENERGIES, UNITS
\newcommand{\nineH}        {$\sqrt{s}~=~0.9$~Te\kern-.1emV\xspace}
\newcommand{\seven}        {$\sqrt{s}~=~7$~Te\kern-.1emV\xspace}
\newcommand{\twoH}         {$\sqrt{s}~=~0.2$~Te\kern-.1emV\xspace}
\newcommand{\twosevensix}  {$\sqrt{s}~=~2.76$~Te\kern-.1emV\xspace}
\newcommand{\five}         {$\sqrt{s}~=~5.02$~Te\kern-.1emV\xspace}
\newcommand{\twosevensixnn}{$\sqrt{s_{\mathrm{NN}}}~=~2.76$~Te\kern-.1emV\xspace}
\newcommand{\fivenn}       {$\sqrt{s_{\mathrm{NN}}}~=~5.02$~Te\kern-.1emV\xspace}
\newcommand{\LT}           {L{\'e}vy-Tsallis\xspace}
\newcommand{\GeVc}         {Ge\kern-.1emV/$c$\xspace}
\newcommand{\MeVc}         {Me\kern-.1emV/$c$\xspace}
\newcommand{\TeV}          {Te\kern-.1emV\xspace}
\newcommand{\GeV}          {Ge\kern-.1emV\xspace}
\newcommand{\MeV}          {Me\kern-.1emV\xspace}
\newcommand{\GeVmass}      {Ge\kern-.2emV/$c^2$\xspace}
\newcommand{\MeVmass}      {Me\kern-.2emV/$c^2$\xspace}
\newcommand{\lumi}         {\ensuremath{\mathcal{L}}\xspace}

% 4) DETECTORS 
\newcommand{\ITS}          {\rm{ITS}\xspace}
\newcommand{\TOF}          {\rm{TOF}\xspace}
\newcommand{\ZDC}          {\rm{ZDC}\xspace}
\newcommand{\ZDCs}         {\rm{ZDCs}\xspace}
\newcommand{\ZNA}          {\rm{ZNA}\xspace}
\newcommand{\ZNC}          {\rm{ZNC}\xspace}
\newcommand{\SPD}          {\rm{SPD}\xspace}
\newcommand{\SDD}          {\rm{SDD}\xspace}
\newcommand{\SSD}          {\rm{SSD}\xspace}
\newcommand{\TPC}          {\rm{TPC}\xspace}
\newcommand{\TRD}          {\rm{TRD}\xspace}
\newcommand{\TZERO}        {\rm{T0}\xspace}
\newcommand{\TZEROA}       {\rm{T0A}\xspace}
\newcommand{\TZEROC}       {\rm{T0C}\xspace}
\newcommand{\VZERO}        {\rm{V0}\xspace}
\newcommand{\VZEROA}       {\rm{V0A}\xspace}
\newcommand{\VZEROC}       {\rm{V0C}\xspace}
\newcommand{\Vdecay} 	   {\ensuremath{\rm V^{0}}\xspace}
% 4) PARTICLE SPECIES 
\newcommand{\ee}           {\ensuremath{e^{+}e^{-}}} 
\newcommand{\pip}          {\ensuremath{\pi^{+}}\xspace}
\newcommand{\pim}          {\ensuremath{\pi^{-}}\xspace}
\newcommand{\kap}          {\ensuremath{\rm{K}^{+}}\xspace}
\newcommand{\kam}          {\ensuremath{\rm{K}^{-}}\xspace}
\newcommand{\pbar}         {\ensuremath{\rm\overline{p}}\xspace}
\newcommand{\kzero}        {\ensuremath{{\rm K}^{0}_{\rm{S}}}\xspace}
\newcommand{\lmb}          {\ensuremath{\Lambda}\xspace}
\newcommand{\almb}         {\ensuremath{\overline{\Lambda}}\xspace}
\newcommand{\Om}           {\ensuremath{\Omega^-}\xspace}
\newcommand{\Mo}           {\ensuremath{\overline{\Omega}^+}\xspace}
\newcommand{\X}            {\ensuremath{\Xi^-}\xspace}
\newcommand{\Ix}           {\ensuremath{\overline{\Xi}^+}\xspace}
\newcommand{\Xis}          {\ensuremath{\Xi^{\pm}}\xspace}
\newcommand{\Oms}          {\ensuremath{\Omega^{\pm}}\xspace}
\newcommand{\degree}       {\ensuremath{^{\rm o}}\xspace}

\newcommand{\Angantyr}       {\rm{PYTHIA8/Angantyr}\xspace}
\newcommand{\Pythia}       {\rm{PYTHIA}\xspace}

%% file: fa_2022-05-09_Opt_C.tex
% Version: 2022-05-09

The ALICE Collaboration would like to thank all its engineers and technicians for their invaluable contributions to the construction of the experiment and the CERN accelerator teams for the outstanding performance of the LHC complex.
The ALICE Collaboration gratefully acknowledges the resources and support provided by all Grid centres and the Worldwide LHC Computing Grid (WLCG) collaboration.
The ALICE Collaboration acknowledges the following funding agencies for their support in building and running the ALICE detector:
A. I. Alikhanyan National Science Laboratory (Yerevan Physics Institute) Foundation (ANSL), State Committee of Science and World Federation of Scientists (WFS), Armenia;
Austrian Academy of Sciences, Austrian Science Fund (FWF): [M 2467-N36] and Nationalstiftung f\"{u}r Forschung, Technologie und Entwicklung, Austria;
Ministry of Communications and High Technologies, National Nuclear Research Center, Azerbaijan;
Conselho Nacional de Desenvolvimento Cient\'{\i}fico e Tecnol\'{o}gico (CNPq), Financiadora de Estudos e Projetos (Finep), Funda\c{c}\~{a}o de Amparo \`{a} Pesquisa do Estado de S\~{a}o Paulo (FAPESP) and Universidade Federal do Rio Grande do Sul (UFRGS), Brazil;
Bulgarian Ministry of Education and Science, within the National Roadmap for Research Infrastructures 2020¿2027 (object CERN), Bulgaria;
Ministry of Education of China (MOEC) , Ministry of Science \& Technology of China (MSTC) and National Natural Science Foundation of China (NSFC), China;
Ministry of Science and Education and Croatian Science Foundation, Croatia;
Centro de Aplicaciones Tecnol\'{o}gicas y Desarrollo Nuclear (CEADEN), Cubaenerg\'{\i}a, Cuba;
Ministry of Education, Youth and Sports of the Czech Republic, Czech Republic;
The Danish Council for Independent Research | Natural Sciences, the VILLUM FONDEN and Danish National Research Foundation (DNRF), Denmark;
Helsinki Institute of Physics (HIP), Finland;
Commissariat \`{a} l'Energie Atomique (CEA) and Institut National de Physique Nucl\'{e}aire et de Physique des Particules (IN2P3) and Centre National de la Recherche Scientifique (CNRS), France;
Bundesministerium f\"{u}r Bildung und Forschung (BMBF) and GSI Helmholtzzentrum f\"{u}r Schwerionenforschung GmbH, Germany;
General Secretariat for Research and Technology, Ministry of Education, Research and Religions, Greece;
National Research, Development and Innovation Office, Hungary;
Department of Atomic Energy Government of India (DAE), Department of Science and Technology, Government of India (DST), University Grants Commission, Government of India (UGC) and Council of Scientific and Industrial Research (CSIR), India;
National Research and Innovation Agency - BRIN, Indonesia;
Istituto Nazionale di Fisica Nucleare (INFN), Italy;
Japanese Ministry of Education, Culture, Sports, Science and Technology (MEXT) and Japan Society for the Promotion of Science (JSPS) KAKENHI, Japan;
Consejo Nacional de Ciencia (CONACYT) y Tecnolog\'{i}a, through Fondo de Cooperaci\'{o}n Internacional en Ciencia y Tecnolog\'{i}a (FONCICYT) and Direcci\'{o}n General de Asuntos del Personal Academico (DGAPA), Mexico;
Nederlandse Organisatie voor Wetenschappelijk Onderzoek (NWO), Netherlands;
The Research Council of Norway, Norway;
Commission on Science and Technology for Sustainable Development in the South (COMSATS), Pakistan;
Pontificia Universidad Cat\'{o}lica del Per\'{u}, Peru;
Ministry of Education and Science, National Science Centre and WUT ID-UB, Poland;
Korea Institute of Science and Technology Information and National Research Foundation of Korea (NRF), Republic of Korea;
Ministry of Education and Scientific Research, Institute of Atomic Physics, Ministry of Research and Innovation and Institute of Atomic Physics and University Politehnica of Bucharest, Romania;
Ministry of Education, Science, Research and Sport of the Slovak Republic, Slovakia;
National Research Foundation of South Africa, South Africa;
Swedish Research Council (VR) and Knut \& Alice Wallenberg Foundation (KAW), Sweden;
European Organization for Nuclear Research, Switzerland;
Suranaree University of Technology (SUT), National Science and Technology Development Agency (NSTDA), Thailand Science Research and Innovation (TSRI) and National Science, Research and Innovation Fund (NSRF), Thailand;
Turkish Energy, Nuclear and Mineral Research Agency (TENMAK), Turkey;
National Academy of  Sciences of Ukraine, Ukraine;
Science and Technology Facilities Council (STFC), United Kingdom;
National Science Foundation of the United States of America (NSF) and United States Department of Energy, Office of Nuclear Physics (DOE NP), United States of America.
In addition, individual groups or members have received support from:
Marie Sk\l{}odowska Curie, European Research Council, Strong 2020 - Horizon 2020 (grant nos. 950692, 824093, 896850), European Union;
Academy of Finland (Center of Excellence in Quark Matter) (grant nos. 346327, 346328), Finland;
Programa de Apoyos para la Superaci\'{o}n del Personal Acad\'{e}mico, UNAM, Mexico.

%% file: 2022-05-09-Alice_Authorlist_2022-05-09_0_Opt_C.tex
% ALICE Collaboration author list for 2022-05-09
\begin{flushleft} 
\small

S.~Acharya\,\orcidlink{0000-0002-9213-5329}\,$^{\rm 125,132}$, 
D.~Adamov\'{a}\,\orcidlink{0000-0002-0504-7428}\,$^{\rm 86}$, 
A.~Adler$^{\rm 69}$, 
G.~Aglieri Rinella\,\orcidlink{0000-0002-9611-3696}\,$^{\rm 32}$, 
M.~Agnello\,\orcidlink{0000-0002-0760-5075}\,$^{\rm 29}$, 
N.~Agrawal\,\orcidlink{0000-0003-0348-9836}\,$^{\rm 50}$, 
Z.~Ahammed\,\orcidlink{0000-0001-5241-7412}\,$^{\rm 132}$, 
S.~Ahmad\,\orcidlink{0000-0003-0497-5705}\,$^{\rm 15}$, 
S.U.~Ahn\,\orcidlink{0000-0001-8847-489X}\,$^{\rm 70}$, 
I.~Ahuja\,\orcidlink{0000-0002-4417-1392}\,$^{\rm 37}$
% ...

\section*{Affiliation Notes}

$^{\rm I}$ Deceased\\
$^{\rm II}$ Also at: Italian National Agency for New Technologies, Energy and Sustainable Economic Development (ENEA), Bologna, Italy\\
$^{\rm III}$ Also at: Dipartimento DET del Politecnico di Torino, Turin, Italy\\
$^{\rm IV}$ Also at: Department of Applied Physics, Aligarh Muslim University, Aligarh, India\\
$^{\rm V}$ Also at: Institute of Theoretical Physics, University of Wroclaw, Poland\\
$^{\rm VI}$ Also at: University of Kansas, Lawrence, Kansas, United States\\
$^{\rm VII}$ Also at: An institution covered by a cooperation agreement with CERN\\

\section*{Collaboration Institutes}

$^{1}$ A.I. Alikhanyan National Science Laboratory (Yerevan Physics Institute) Foundation, Yerevan, Armenia\\
$^{2}$ AGH University of Science and Technology, Cracow, Poland\\
$^{3}$ Bogolyubov Institute for Theoretical Physics, National Academy of Sciences of Ukraine, Kiev, Ukraine\\
$^{4}$ Bose Institute, Department of Physics  and Centre for Astroparticle Physics and Space Science (CAPSS), Kolkata, India\\
$^{5}$ California Polytechnic State University, San Luis Obispo, California, United States\\
$^{6}$ Central China Normal University, Wuhan, China\\
$^{7}$ Centro de Aplicaciones Tecnol\'{o}gicas y Desarrollo Nuclear (CEADEN), Havana, Cuba\\
$^{8}$ Centro de Investigaci\'{o}n y de Estudios Avanzados (CINVESTAV), Mexico City and M\'{e}rida, Mexico\\
$^{9}$ Chicago State University, Chicago, Illinois, United States\\
$^{10}$ China Institute of Atomic Energy, Beijing, China\\
$^{11}$ Chungbuk National University, Cheongju, Republic of Korea\\
$^{12}$ Comenius University Bratislava, Faculty of Mathematics, Physics and Informatics, Bratislava, Slovak Republic\\
$^{13}$ COMSATS University Islamabad, Islamabad, Pakistan\\
$^{14}$ Creighton University, Omaha, Nebraska, United States\\
$^{15}$ Department of Physics, Aligarh Muslim University, Aligarh, India\\
$^{16}$ Department of Physics, Pusan National University, Pusan, Republic of Korea\\
$^{17}$ Department of Physics, Sejong University, Seoul, Republic of Korea\\
$^{18}$ Department of Physics, University of California, Berkeley, California, United States\\
$^{19}$ Department of Physics, University of Oslo, Oslo, Norway\\
$^{20}$ Department of Physics and Technology, University of Bergen, Bergen, Norway\\
$^{21}$ Dipartimento di Fisica, Universit\`{a} di Pavia, Pavia, Italy\\
$^{22}$ Dipartimento di Fisica dell'Universit\`{a} and Sezione INFN, Cagliari, Italy\\
$^{23}$ Dipartimento di Fisica dell'Universit\`{a} and Sezione INFN, Trieste, Italy\\
$^{24}$ Dipartimento di Fisica dell'Universit\`{a} and Sezione INFN, Turin, Italy\\
$^{25}$ Dipartimento di Fisica e Astronomia dell'Universit\`{a} and Sezione INFN, Bologna, Italy\\
$^{26}$ Dipartimento di Fisica e Astronomia dell'Universit\`{a} and Sezione INFN, Catania, Italy\\
$^{27}$ Dipartimento di Fisica e Astronomia dell'Universit\`{a} and Sezione INFN, Padova, Italy\\
$^{28}$ Dipartimento di Fisica `E.R.~Caianiello' dell'Universit\`{a} and Gruppo Collegato INFN, Salerno, Italy\\
$^{29}$ Dipartimento DISAT del Politecnico and Sezione INFN, Turin, Italy\\
$^{30}$ Dipartimento di Scienze MIFT, Universit\`{a} di Messina, Messina, Italy\\
$^{31}$ Dipartimento Interateneo di Fisica `M.~Merlin' and Sezione INFN, Bari, Italy\\
$^{32}$ European Organization for Nuclear Research (CERN), Geneva, Switzerland\\
$^{33}$ Faculty of Electrical Engineering, Mechanical Engineering and Naval Architecture, University of Split, Split, Croatia\\
$^{34}$ Faculty of Engineering and Science, Western Norway University of Applied Sciences, Bergen, Norway\\
$^{35}$ Faculty of Nuclear Sciences and Physical Engineering, Czech Technical University in Prague, Prague, Czech Republic\\
$^{36}$ Faculty of Physics, Sofia University, Sofia, Bulgaria\\
$^{37}$ Faculty of Science, P.J.~\v{S}af\'{a}rik University, Ko\v{s}ice, Slovak Republic\\
$^{38}$ Frankfurt Institute for Advanced Studies, Johann Wolfgang Goethe-Universit\"{a}t Frankfurt, Frankfurt, Germany\\
$^{39}$ Fudan University, Shanghai, China\\
$^{40}$ Gangneung-Wonju National University, Gangneung, Republic of Korea\\
$^{41}$ Gauhati University, Department of Physics, Guwahati, India\\
$^{42}$ Helmholtz-Institut f\"{u}r Strahlen- und Kernphysik, Rheinische Friedrich-Wilhelms-Universit\"{a}t Bonn, Bonn, Germany\\
$^{43}$ Helsinki Institute of Physics (HIP), Helsinki, Finland\\
$^{44}$ High Energy Physics Group,  Universidad Aut\'{o}noma de Puebla, Puebla, Mexico\\
$^{45}$ Horia Hulubei National Institute of Physics and Nuclear Engineering, Bucharest, Romania\\
$^{46}$ Indian Institute of Technology Bombay (IIT), Mumbai, India\\
$^{47}$ Indian Institute of Technology Indore, Indore, India\\
$^{48}$ INFN, Laboratori Nazionali di Frascati, Frascati, Italy\\
$^{49}$ INFN, Sezione di Bari, Bari, Italy\\
$^{50}$ INFN, Sezione di Bologna, Bologna, Italy\\
$^{51}$ INFN, Sezione di Cagliari, Cagliari, Italy\\
$^{52}$ INFN, Sezione di Catania, Catania, Italy\\
$^{53}$ INFN, Sezione di Padova, Padova, Italy\\
$^{54}$ INFN, Sezione di Pavia, Pavia, Italy\\
$^{55}$ INFN, Sezione di Torino, Turin, Italy\\
$^{56}$ INFN, Sezione di Trieste, Trieste, Italy\\
$^{57}$ Inha University, Incheon, Republic of Korea\\
$^{58}$ Institute for Gravitational and Subatomic Physics (GRASP), Utrecht University/Nikhef, Utrecht, Netherlands\\
$^{59}$ Institute of Experimental Physics, Slovak Academy of Sciences, Ko\v{s}ice, Slovak Republic\\
$^{60}$ Institute of Physics, Homi Bhabha National Institute, Bhubaneswar, India\\
$^{61}$ Institute of Physics of the Czech Academy of Sciences, Prague, Czech Republic\\
$^{62}$ Institute of Space Science (ISS), Bucharest, Romania\\
$^{63}$ Institut f\"{u}r Kernphysik, Johann Wolfgang Goethe-Universit\"{a}t Frankfurt, Frankfurt, Germany\\
$^{64}$ Instituto de Ciencias Nucleares, Universidad Nacional Aut\'{o}noma de M\'{e}xico, Mexico City, Mexico\\
$^{65}$ Instituto de F\'{i}sica, Universidade Federal do Rio Grande do Sul (UFRGS), Porto Alegre, Brazil\\
$^{66}$ Instituto de F\'{\i}sica, Universidad Nacional Aut\'{o}noma de M\'{e}xico, Mexico City, Mexico\\
$^{67}$ iThemba LABS, National Research Foundation, Somerset West, South Africa\\
$^{68}$ Jeonbuk National University, Jeonju, Republic of Korea\\
$^{69}$ Johann-Wolfgang-Goethe Universit\"{a}t Frankfurt Institut f\"{u}r Informatik, Fachbereich Informatik und Mathematik, Frankfurt, Germany\\
$^{70}$ Korea Institute of Science and Technology Information, Daejeon, Republic of Korea\\
$^{71}$ KTO Karatay University, Konya, Turkey\\
$^{72}$ Laboratoire de Physique des 2 Infinis, Ir\`{e}ne Joliot-Curie, Orsay, France\\
$^{73}$ Laboratoire de Physique Subatomique et de Cosmologie, Universit\'{e} Grenoble-Alpes, CNRS-IN2P3, Grenoble, France\\
$^{74}$ Lawrence Berkeley National Laboratory, Berkeley, California, United States\\
$^{75}$ Lund University Department of Physics, Division of Particle Physics, Lund, Sweden\\
$^{76}$ Nagasaki Institute of Applied Science, Nagasaki, Japan\\
$^{77}$ Nara Women{'}s University (NWU), Nara, Japan\\
$^{78}$ National and Kapodistrian University of Athens, School of Science, Department of Physics , Athens, Greece\\
$^{79}$ National Centre for Nuclear Research, Warsaw, Poland\\
$^{80}$ National Institute of Science Education and Research, Homi Bhabha National Institute, Jatni, India\\
$^{81}$ National Nuclear Research Center, Baku, Azerbaijan\\
$^{82}$ National Research and Innovation Agency - BRIN, Jakarta, Indonesia\\
$^{83}$ Niels Bohr Institute, University of Copenhagen, Copenhagen, Denmark\\
$^{84}$ Nikhef, National institute for subatomic physics, Amsterdam, Netherlands\\
$^{85}$ Nuclear Physics Group, STFC Daresbury Laboratory, Daresbury, United Kingdom\\
$^{86}$ Nuclear Physics Institute of the Czech Academy of Sciences, Husinec-\v{R}e\v{z}, Czech Republic\\
$^{87}$ Oak Ridge National Laboratory, Oak Ridge, Tennessee, United States\\
$^{88}$ Ohio State University, Columbus, Ohio, United States\\
$^{89}$ Physics department, Faculty of science, University of Zagreb, Zagreb, Croatia\\
$^{90}$ Physics Department, Panjab University, Chandigarh, India\\
$^{91}$ Physics Department, University of Jammu, Jammu, India\\
$^{92}$ Physics Department, University of Rajasthan, Jaipur, India\\
$^{93}$ Physics Program and International Institute for Sustainability with Knotted Chiral Meta Matter (SKCM2), Hiroshima University, Hiroshima, Japan\\
$^{94}$ Physikalisches Institut, Eberhard-Karls-Universit\"{a}t T\"{u}bingen, T\"{u}bingen, Germany\\
$^{95}$ Physikalisches Institut, Ruprecht-Karls-Universit\"{a}t Heidelberg, Heidelberg, Germany\\
$^{96}$ Physik Department, Technische Universit\"{a}t M\"{u}nchen, Munich, Germany\\
$^{97}$ Politecnico di Bari and Sezione INFN, Bari, Italy\\
$^{98}$ Research Division and ExtreMe Matter Institute EMMI, GSI Helmholtzzentrum f\"ur Schwerionenforschung GmbH, Darmstadt, Germany\\
$^{99}$ Saga University, Saga, Japan\\
$^{100}$ Saha Institute of Nuclear Physics, Homi Bhabha National Institute, Kolkata, India\\
$^{101}$ School of Physics and Astronomy, University of Birmingham, Birmingham, United Kingdom\\
$^{102}$ Secci\'{o}n F\'{\i}sica, Departamento de Ciencias, Pontificia Universidad Cat\'{o}lica del Per\'{u}, Lima, Peru\\
$^{103}$ Stefan Meyer Institut f\"{u}r Subatomare Physik (SMI), Vienna, Austria\\
$^{104}$ SUBATECH, IMT Atlantique, Nantes Universit\'{e}, CNRS-IN2P3, Nantes, France\\
$^{105}$ Suranaree University of Technology, Nakhon Ratchasima, Thailand\\
$^{106}$ Technical University of Ko\v{s}ice, Ko\v{s}ice, Slovak Republic\\
$^{107}$ The Henryk Niewodniczanski Institute of Nuclear Physics, Polish Academy of Sciences, Cracow, Poland\\
$^{108}$ The University of Texas at Austin, Austin, Texas, United States\\
$^{109}$ Universidad Aut\'{o}noma de Sinaloa, Culiac\'{a}n, Mexico\\
$^{110}$ Universidade de S\~{a}o Paulo (USP), S\~{a}o Paulo, Brazil\\
$^{111}$ Universidade Estadual de Campinas (UNICAMP), Campinas, Brazil\\
$^{112}$ Universidade Federal do ABC, Santo Andre, Brazil\\
$^{113}$ University of Cape Town, Cape Town, South Africa\\
$^{114}$ University of Houston, Houston, Texas, United States\\
$^{115}$ University of Jyv\"{a}skyl\"{a}, Jyv\"{a}skyl\"{a}, Finland\\
$^{116}$ University of Kansas, Lawrence, Kansas, United States\\
$^{117}$ University of Liverpool, Liverpool, United Kingdom\\
$^{118}$ University of Science and Technology of China, Hefei, China\\
$^{119}$ University of South-Eastern Norway, Kongsberg, Norway\\
$^{120}$ University of Tennessee, Knoxville, Tennessee, United States\\
$^{121}$ University of the Witwatersrand, Johannesburg, South Africa\\
$^{122}$ University of Tokyo, Tokyo, Japan\\
$^{123}$ University of Tsukuba, Tsukuba, Japan\\
$^{124}$ University Politehnica of Bucharest, Bucharest, Romania\\
$^{125}$ Universit\'{e} Clermont Auvergne, CNRS/IN2P3, LPC, Clermont-Ferrand, France\\
$^{126}$ Universit\'{e} de Lyon, CNRS/IN2P3, Institut de Physique des 2 Infinis de Lyon, Lyon, France\\
$^{127}$ Universit\'{e} de Strasbourg, CNRS, IPHC UMR 7178, F-67000 Strasbourg, France, Strasbourg, France\\
$^{128}$ Universit\'{e} Paris-Saclay Centre d'Etudes de Saclay (CEA), IRFU, D\'{e}partment de Physique Nucl\'{e}aire (DPhN), Saclay, France\\
$^{129}$ Universit\`{a} degli Studi di Foggia, Foggia, Italy\\
$^{130}$ Universit\`{a} del Piemonte Orientale, Vercelli, Italy\\
$^{131}$ Universit\`{a} di Brescia, Brescia, Italy\\
$^{132}$ Variable Energy Cyclotron Centre, Homi Bhabha National Institute, Kolkata, India\\
$^{133}$ Warsaw University of Technology, Warsaw, Poland\\
$^{134}$ Wayne State University, Detroit, Michigan, United States\\
$^{135}$ Westf\"{a}lische Wilhelms-Universit\"{a}t M\"{u}nster, Institut f\"{u}r Kernphysik, M\"{u}nster, Germany\\
$^{136}$ Wigner Research Centre for Physics, Budapest, Hungary\\
$^{137}$ Yale University, New Haven, Connecticut, United States\\
$^{138}$ Yonsei University, Seoul, Republic of Korea\\
$^{139}$  Zentrum  f\"{u}r Technologie und Transfer (ZTT), Worms, Germany\\
$^{140}$ Affiliated with an institute covered by a cooperation agreement with CERN\\
$^{141}$ Affiliated with an international laboratory covered by a cooperation agreement with CERN\\

\end{flushleft} 